\documentclass[twocolumn,showpacs,preprintnumbers,amsmath,amssymb,superscriptaddress]{revtex4}

\usepackage{longtable}
\usepackage{graphicx}
\usepackage{dcolumn}
\usepackage{bm}

\begin{document}

\title{Muon-Induced Background Study for an Argon-Based Long Baseline Neutrino Experiment}

\newcommand{\usd}{Department of Physics, The University of South Dakota, Vermillion, South Dakota 57069}
\newcommand{\tgu}{College of Sciences, China Three Gorges University, Yichang 443002, China}
\newcommand{\dmm}{Dongming.Mei@usd.edu}

\author{ D. Barker} \affiliation{ \usd}
\author{     D.-M.~Mei          }\altaffiliation[Corresponding Author: ]{\dmm}	\affiliation{	\usd 	}
\author{    C. Zhang           }\affiliation{ \usd} \affiliation{\tgu} 			

\begin{abstract}

We evaluated rates of transversing muons, muon-induced fast neutrons, and production of $^{40}$Cl and other 
cosmogenically produced nuclei that pose as potential sources of background to the physics program proposed for an argon-based long baseline neutrino
experiment at the Sanford Underground Research Facility (SURF), Homestake mine, Lead, SD. The Geant4 simulations were carried out with muons and
muon-induced neutrons for both the 800-ft level (0.712 km.w.e.) and 4850-ft level (4.3 km.w.e.).  
We developed analytic models to independently calculate the $^{40}$Cl production using the
measured muon fluxes at different levels of the Homestake mine. The muon induced $^{40}$Cl production rates through stopped muon capture and 
the muon-induced neutrons and protons via
(n,p) and (p,n) reactions were evaluated. We find that the Monte Carlo simulated production rates of $^{40}$Cl agree well
with the predictions from analytic models. A depth-dependent parametrization was developed and benchmarked to the direct analytic models. 
 We conclude that the muon-induced processes will result in large
backgrounds to the physics proposed for an argon-based long baseline neutrino experiment at a
depth of less than 4.0 km.w.e.
  
\end{abstract}

\pacs{13.85.Hd, 23.40.-s, 25.40.Fq}
\maketitle
\section{Introduction}
Experiments for the past several decades require modification of the Standard Model to incorporate the unexpected neutrino properties and fundamental
 characteristics~\cite{superk, macro, soudanii, k2k, minos, sno, homestake, kamland}. For instance, neutrino flavor mixing was found to be responsible for the phenomenon
of neutrino oscillation~\cite{superk, macro, soudanii, k2k, minos, sno, homestake, kamland} occurring between three generations in which a 
complex phase ($\delta_{CP}$) 
signifies the amount of violation of the charge-parity (CP) symmetry, which is unknown. The sign of the mass difference, 
$\Delta m_{13}^2$, which represents the ordering of the mass eigenstates, remains unknown as well. 
These two unknown parameters are intended to be addressed with the next generation of neutrino oscillation experiments. 
The value of the mixing angle, $\theta_{13}$, between the first generation and the third generation, was recently reported by the Double Chooz~\cite{doublechooz2}, 
Daya Bay~\cite{dayabay} 
and Reno~\cite{reno} collaborations to be large ($\sim$9$^{o}$). Neutrino beam experiments can provide an independent check to verify 
the results reported by these reactor experiments.

Recent studies of neutrino properties using neutrino beams have demonstrated that good sensitivity to CP violation and mass hierarchy can be achieved 
by measuring $\nu_{e}$ appearance using a very long baseline $\nu_{\mu}$ beam with massive detectors~\cite{Barger} assuming the value of $\theta_{13}$ $>$ 1$^{o}$.
The recent measurements of $\theta_{13}$ made by MINOS and T2K show a non-zero value~\cite{minos1, t2k}. The most recent values of $\theta_{13}$ reported by reactor 
experiments show a big value of around 9$^{o}$ that is particularly interesting to the measurements of 
CP violation and
mass hierarchy. The proof of CP violation in the lepton sector and the knowledge
of the value of $\delta_{CP}$ are crucial to understanding the origin of the baryon
asymmetry in the universe, providing a strong implication of leptogenesis that is responsible for the observed baryon asymmetry of the universe~\cite{FMY}. 
Simultaneously, the neutrino mass hierarchy is of great importance for neutrinoless double-beta decay experiments~\cite{FFS} and could shed light on
possible flavor symmetries. 

When measuring the value of $\theta_{13}$, CP phase, $\delta_{CP}$, and the neutrino mass hierarchy with conventional neutrino beams, a key process of new discovery 
in neutrino oscillation is $\nu_{\mu} \rightarrow \nu_{e}$ appearance. With appropriate detector design and adequate control of environmental factors, specifically proper 
shielding, the long baseline neutrino experiment (LBNE) ~\cite{lbnecdr, LBNE} is capable of supporting an extremely rich program of high energy physics and particle-astrophysics 
including proton decay, astronomical neutrinos, and tests of fundamental physics and the Standard Model. 
The two detector technologies being considered are: 1) an active finely grained liquid argon time-projection-chamber
 (LAr-TPC) and 2) a water Cerenkov detector. Both detector technologies can support this wide range of physics goals probing the Standard Model and 
searching for physics beyond the current models. However, the muon-induced background may constrain the sensitivity 
of the proposed experiment, in particular, the proposed galactic and relic supernova neutrino detection with a liquid argon detector. Though the background 
can be measured for beam neutrino physics with the beam-on or -off, the fluctuation of background events can be problematic if these events occur 
on the same order of magnitude
as the anticipated physics signal.  In addition to the beam contamination backgrounds, the main sources are the muon-induced processes. 
In this paper, we report the
study of the muon-induced background for an argon-based detector.

Muons and muon-induced fast neutrons entering the detector from the surrounding rock can cause troublesome experimental backgrounds. 
While through-going muon events 
in the detector have energy depositions and can be largely identified, a small fraction of the muon-induced energetic delta electrons and muons that 
traverse the rock near the detector, or traverse very small distances in the detector ("corner-clippers"), can result in limiting backgrounds. Moreover, 
background events 
can be produced by the 
 muon-induced fast neutrons entering the detector while the parent muons pass completely through the surrounding in-active materials.   
The muon-induced neutrons can undergo (n,p) reaction with $^{40}$Ar, and negative
 muons capture on $^{40}$Ar.  Both reactions create unwanted $^{40}$Cl, which can be a background for
the following reactions: $\bar{\nu_{e}} + ^{40}$Ar $\rightarrow e^{+} + ^{40}$Cl, $\nu_{e} + ^{40}$Ar$ \rightarrow e^{-} + ^{40}$K, 
$\nu_{x} + ^{40}$Ar $\rightarrow$ $\nu_{x} + ^{40}$Ar$^{*}$, and $\nu_{x,\bar{x}} + e^{-}$ $\rightarrow$ $\nu_{x} + e^{-}$, where $x$ = $e$, $\mu$, $\tau$. 
This is  because the decay
Q value of $^{40}$Cl is 7.48 MeV, which is above the proposed detection threshold of 5 MeV or 6 MeV, and the half-life of $^{40}$Cl is 1.35 minutes, 
making it difficult to correlate with muons.
There can also be additional radioactive isotopes produced 
by muon-induced processes in the argon
target. Those radioactive isotopes can be part of the background for the proposed physics channels. 
Therefore, it is necessary to evaluate the muon-induced processes and the production rate of the cosmogenics in the detector volume.

Due to the lack of direct detection measurements of the muons and muon-induced products and parameters for a given depth with a given detector, 
such an evaluation has not been modelled in full. In this paper, we present several parametrization functions that estimate the muon-induced fast neutron 
energy spectrum and the stopping muons as a function of depth. 
Using these parametrization functions, we simulated cosmogenic production rates in the proposed LBNE LAr detector 
for two depths with a well known Geant4 package~\cite{geant4}, Geant4.9.5 with shielding physics list. The simulated results were compared to the predictions by the developed analytic models. 
Since the  muon-induced processes are strongly depth dependent~\cite{meihime}, 
we establish a depth-sensitivity 
relation for an argon-based detector by calculating the production rate of cosmogenics as a function of depth.  

\section{Evaluation of Muon-Induced Background}
At sufficiently high energies, radiative processes become predominant in energy loss for muons. Many sub-sequential particles can be produced by muon-induced
radiative processes. The energetic delta electrons induced by muons in the ionization process can also undergo bremsstrahlung radiation. 
Cosmogenic radioactive isotopes can be produced by muons and muon-induced sequential particles including neutrons, protons, pions, gamma rays, etc. Two cases are
considered in the following evaluation: 1) muons transversing the detector and 2) muon-induced neutrons entering the detector from the experimental hall. 
The former process creates $^{40}$Cl and other radioactive isotopes through negative muon capture, (n,p) reaction, and (p,n) reaction, etc. The latter produces mainly $^{40}$Cl via (n,p) reaction alone. In order to understand the production rates and their corresponding mechanisms, we performed a full Geant4 Monte Carlo simulation and
developed analytic models. We elaborate on the evaluation processes in the following subsections. 

\subsection{Muon-induced Background from the Geant4 Simulation}
High energy cosmic-ray muons can penetrate rock overburden to reach an underground laboratory with the surviving muons generating neutrons in the surrounding rock. 
Those neutrons are unwanted particles that could produce background events for low-background
experiments searching for rare event physics. The intensities of the residual muons and the muon-induced neutrons depend strongly on the depth of the underground
detector. This is particularly important for detecting supernova neutrinos that contribute to a signal range of a few MeV to a few tens of MeV, which 
can often be dominated by the muon-induced backgrounds depending on the depth. Therefore, the depth-sensitivity relation needs to be understood in order 
to choose an appropriate depth at the Sanford Underground Research Facility (SURF) to estimate the experimental limitations imposed by these backgrounds 
and to optimize the detector design for the full range of possible physics programs. Because both the 800-ft level and the 4850-ft level are being considered
 for a far detector, we have conducted Geant4 Monte Carlo simulations
to understand the cosmogenic production in the detector at both levels. 

In the Geant4 simulation, a simple geometry with a dimension of liquid argon, 45.6 m (width) $\times$ 22.4 m (length) $\times$ 14.0 m (height)~\cite{lbnecdr}, 
was placed in a 
stainless steel container with a thickness of 1 cm. Both muons and  neutrons were generated on a very thin sheet of air (100 m (width) $\times$ 50 m (length)
 $\times$0.01 m (thickness)) right above the upper
surface of the stainless steel tank. Since the thin sheet is two times larger than the size of the detector in length and width, both muons and neutrons generated
 in the sheet
can enter the detector with angular distribution of $sec(\theta)$. There is a wide range of thetas that are enabled by taking a large production sheet 
and better represent the range of anticipated underground muons. Note that we assume the parent muons and daughter neutrons have the 
same angular distribution.
The residual muon energy spectrum was obtained from~\cite{tkg, sei}
\begin{equation}
\label{eq:MeiHime8}
\frac{dN}{dE_{\mu}} = Ae^{-bh(\gamma_{\mu}-1)}\cdot\left(E_{\mu}+\epsilon_{\mu}\left(1-e^{-bh}\right)\right)^{-\gamma_{\mu}}
\end{equation}
where $A$ is a normalization constant determined using the differential muon intensity at a specific depth, $E_{\mu}$ is the muon energy after 
traversing the rock slant depth $h$ (km.w.e), and the parameters are $b=0.4/$km.w.e, $\gamma_{\mu} = 3.77$, and $\epsilon_{\mu} = 693$ GeV~\cite{deg2}.
 
The differential muon intensity used to solve for the normalization constant, $A$, is given by~\cite{meihime}
\begin{eqnarray}
I_{\mu}(h_{0}) = 67.97\times 10^{-6}e^{-h_{0}/0.285} \nonumber \\
+ 2.071\times 10^{-6}e^{-h_{0}/0.698}
\label{eq:MeiHime4}
\end{eqnarray}
with $h_{0}$ the vertical depth (km.w.e). The units of $I_{\mu}(h_{0})$ are cm$^{-2}$s$^{-1}$ which is appropriate in the flat-earth approximation~\cite{meihime}.

The muon-induced neutron energy spectrum is given by the parametrization fitting function~\cite{meihime},
\begin{equation}
\label{eq:MeiHime14}
\frac{dN}{dE_{n}} = A_{\mu}\left(\frac{e^{-a_{0}E_{n}}}{E_{n}} + B_{\mu}(E_{\mu})e^{-a_{1}E_{n}}\right) + a_{2}E_{n}^{-a_{3}}
\end{equation}
where $A_{\mu}$ is a normalization constant, $E_{n}$ is the neutron energy, $a_{0},a_{1},a_{2}$ and $a_{3}$ are fitted parameters, and $B_{\mu}(E_{\mu})$ is 
a function of muon energy with $E_{\mu}$ in GeV,
\begin{equation}
B_{\mu}(E_{\mu}) = 0.324 - 0.641e^{-0.014E_{\mu}}.
\end{equation}
This parametrization is valid for $E_{n} > 10$ MeV and consistent with Ref.~\cite{yfw}.

However, these
equations cannot be directly used to generate neutrons for a given depth without knowing the associated parameters. We adopt the following procedures to
obtain the neutron energy spectrum as a function of depth. First, the average neutron energy as a function of depth was studied
 using the simulated results for various depths in Tables I and VII from Ref.~\cite{meihime} and a measured surface data point from Ref.~\cite{gor}. 
The relevant information is displayed in Table~\ref{tab:AveNeu} and Table~\ref{tab:AveNeu1}. 
\begin{table}[htb!!!]
\caption{The average neutron energy measured on the surface~\cite{gor} and at various underground sites~\cite{meihime}, and the equivalent vertical depth relative to a flat overburden.}
\label{tab:AveNeu}
\begin{tabular}{|l|l|l|}
\hline \hline 
Site & Depth (km.w.e) & $<E_{n}>$ (MeV)\\
\hline
Surface & 0 & 6.5\\
WIPP & 1.585 & 62\\
Soudan & 1.95 & 76\\
Kamioka & 2.05 & 79\\
Boulby & 2.805 & 88\\
Gran Sasso & 3.1 & 91\\
Sudbury & 6.011 & 109\\
\hline \hline
\end{tabular}
\end{table}

\begin{table}[htb!!!]
\caption{Fitting parameters at various underground sites, and the equivalent vertical depth relative to a flat overburden~\cite{meihime}.}
\label{tab:AveNeu1}
\begin{tabular}{|l|l|l|l|l|l|}
\hline \hline 
Site & Depth & $a_{0}$ & $a_{1}$ & $a_{2}$ & $a_{3}$\\
 & (km.w.e)  & & & & \\
\hline
WIPP & 1.585 & 6.86 & 2.1 & 2.971 $\times 10^{-13}$ & 2.456\\
Soudan & 1.95 & 7.333 & 2.105 & -5.35 $\times 10^{-15}$ & 2.893\\
Kamioka & 2.05  & 7.55 & 2.118 & -1.258 $\times 10^{-14}$ & 2.761\\
Boulby & 2.805 & 7.882 & 2.212 & -2.342 $\times 10^{-14}$ & 2.613\\
Gran Sasso & 3.1 & 7.828 & 2.23 & -7.505 $\times 10^{-15}$ & 2.831\\
Sudbury & 6.011  & 7.774 & 2.134& -2.939 $\times 10^{-16}$ & 2.859\\
\hline \hline
\end{tabular}
\end{table}

\begin{figure}[htb!!!]
\includegraphics[angle=0,width=8.cm] {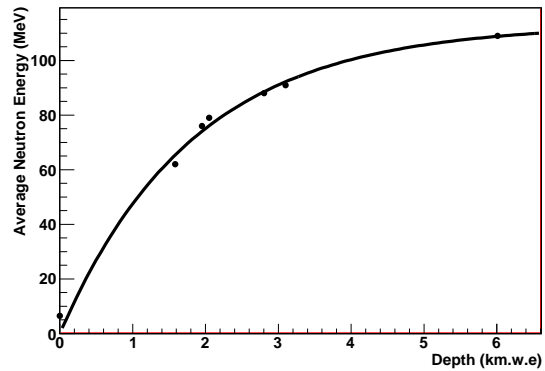}
\caption{\small{Average neutron energy as a function of depth. }}
\label{fig:AveNeu}
\end{figure}

To validate the simulation package, we simulated the surface neutron energy spectrum. The result agrees with the measured spectrum very well.
The main sources of uncertainty in the simulation are the rock density and the distribution of chemical composition as well as the water content 
as a function of depth. In quantifying the uncertainty, we have varied the rock density and chemical composition by 10\%, and the water content 
from dry (8\%) to wet (16\%) in the Monte Carlo simulation. The variation of the average neutron energy is always less than 35\%.  This is to 
say that the neutron energy spectrum as a function of depth obtained from the Monte Carlo simulation is accurate within 35\%. 
Utilizing the average neutron energies obtained for the surface and several depths underground, we obtained a parametrization function of
 the average neutron energy as a function of depth.
Fig.~\ref{fig:AveNeu} shows the fitted curve. 

The fitted function can be expressed as
\begin{equation}
<E_n> = 113\cdot[1-exp(-0.545\cdot h)],
\end{equation}
where $h$ is the depth in km.w.e. At 800-ft, $<E_n>$ is 36.3 MeV, and at 4850-ft, $<E_n>$ is 102.2 MeV. 
It is noticed that the average neutron energy at the depth of Gran Sasso is 
about 91 MeV~\cite{meihime}, which is similar to the average neutron energy predicted for the 4850-ft level at SURF. This is because
the production mechanisms are primarily sensitive to the overburden of rock and do not depend much on site specific details. 
Therefore, 
 we then used the parameters in Table~\ref{tab:AveNeu1} (Table VII of Ref.~\cite{meihime}) to generate neutrons for 
the depth of Gran Sasso. The normalization is done with the following equation from Ref.~\cite{meihime} that predicts the muon-induced neutron 
flux as a function of depth
for depths greater than 1.6 km.w.e:
\begin{equation}
\label{eq:MeiHime13}
\phi_{n} = P_{0}\left(\frac{P_{1}}{h_{0}}\right)exp(-h_{0}/P_{1})
\end{equation}
where, again, $h_{0}$ is the equivalent depth in km.w.e relative to a flat overburden, and the fitting parameters are $P_{0} = (4.0\pm 1.1)\times 10^{-7}$ 
cm$^{-2}$s$^{-1}$ and $P_{1} = 0.86\pm0.05$ km.w.e.
 One caveat in this method is the
ignorance of rock composition, which could make a difference of up to 35\%~\cite{meihime}.
 The muon-induced neutron energy spectrum at the 800-ft level was obtained using a scaling method.   
 We scaled the neutron energy spectrum from the depth of 4850-ft to the depth of the 800-ft 
using a scaling factor, $\frac{<E_{\mu, 800}>\Phi_{\mu,800}}{<E_{\mu,4850}>\Phi_{\mu,4850}}$, where $<E_{\mu,800}>$, $\Phi_{\mu,800}$, and $<E_{\mu,4850}>$, $\Phi_{\mu,4850}$, are
the average muon energies and the total muon fluxes for the 800-ft level and the 4850-ft level at SURF, respectively. 
Note that the primary concerns of the cosmogenic production are $^{40}$Cl and $^{40}$K through negative muon capture, (n,p), and (p,n) reactions. Since the reaction threshold
of $^{40}$Ar(n,p)$^{40}$Cl requires neutrons with kinetic energy greater than 6.87 MeV, this cuts the majority of neutrons induced by natural radioactivity in rock 
through ($\alpha$,n) reactions. Therefore, we neglected
the calculation of $^{   40}$Cl production by ($\alpha$,n) neutrons.

\subsubsection{Muon-induced Background at the 800-ft Level}
The muons that survived the 800-ft rock at SURF have a calculated average energy of 97 GeV and a flux of 6.3$\times$10$^{-6}$cm$^{-2}$s$^{-1}$
 from Eq.\ref{eq:MeiHime4} and the following equation~\cite{meihime} 
\begin{equation}
\label{eq:MeiHime9}
<E_{\mu}> = \frac{\epsilon_{\mu}\left(1-exp(-bh)\right)}{\gamma_{\mu} - 2}
\end{equation}
where $\epsilon_{\mu}=693$ GeV, $b = 0.4$ km.w.e, $h$ is the depth in km.w.e~\cite{deg2}, and $\gamma_{\mu} = 3.77$~\cite{lip}.

\begin{figure}[htb!!!]
\includegraphics[angle=0,width=8.0cm] {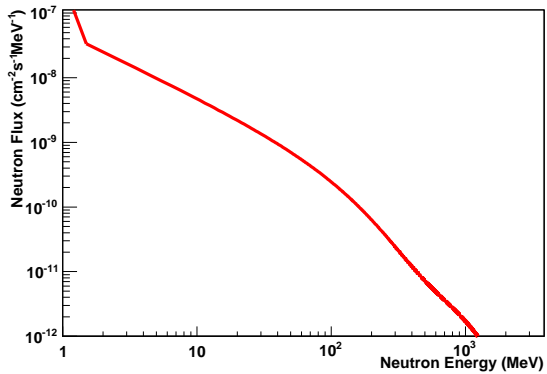}
\caption{\small{
Local fast neutron energy spectrum at the 800-ft level used in the Monte Carlo simulation.}}
\label{fig:neuspe800}
\end{figure}

\begin{figure}[htb!!!]
\includegraphics[angle=0,width=8.0cm] {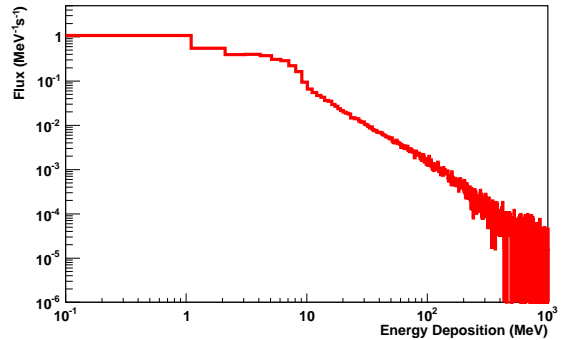}
\caption{\small{
Fast neutrons induced visible energy spectrum in the detector at the 800-ft level. Lindhard's ionization efficiency~\cite{lindhard} was applied to the nuclear recoil events in this plot.}}
\label{fig:neuvisible800}
\end{figure}

These high energy muons passing through the surrounding rock of a laboratory will generate 
fast neutrons. The emerging neutrons in an experimental hall have an average energy of $~$36 MeV with a total flux of 3.2$\times$10$^{-7}$cm$^{-2}$s$^{-1}$ 
from the Monte Carlo simulation. The
neutron energy spectrum obtained from the methods described above is shown in Fig.~\ref{fig:neuspe800}.
We simulated the muons passing through the detector and the muon-induced neutrons entering the detector from rock. 
Fig.~\ref{fig:neuvisible800} shows a visible energy spectrum induced by the muon-induced neutrons entering the detector from the
surrounding rock. The event rate above 5 MeV in a 20 kton detector is $\sim$0.28 Hz. The direct muon rate is $\sim$88 Hz.
There rates are high enough
to potentially swamp any signals from a galactic supernova ($\sim$44 Hz estimated using Ref.~\cite{icarus}) or even from a neutrino beam ($\sim$75 events per year 
for $\nu_{e}$ appearance estimated with Ref.~\cite{beam}). 

\begin{figure}[htb!!!]
\includegraphics[angle=0,width=8.0cm] {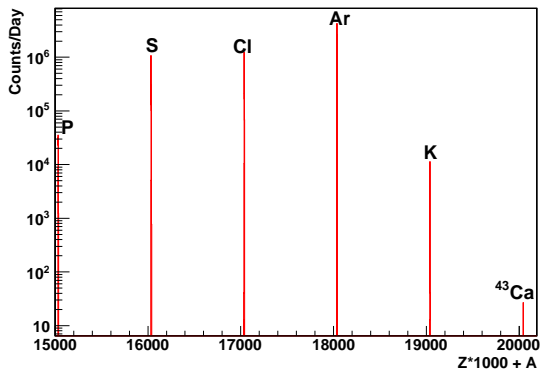}
\caption{\small{
Muon-induced cosmogenic production at the 800-ft level from the Monte Carlo.}}
\label{fig:total800}
\end{figure}

We summarize the $^{40}$Cl production rates from the above two sources in Table~\ref{tab:Sim800ft}. 
\begin{table}[htb!!!]
\caption{ $^{40}$Cl production rates in the detector (20 kton) at the 800-ft level from the Monte Carlo simulation.}
\label{tab:Sim800ft}
\begin{tabular}{|l|l|l|l|}
\hline \hline 
\multicolumn{2}{|c|}{From $\mu$ simulation} & \multicolumn{2}{c|}{From $n$ simulation}\\
\hline
Produced by & Rate per day & Produced by & Rate per day\\
\hline
Muon Capture  & 27344 & Secondary $\mu$ & 45\\
Secondary n & 40587 & Neutrons & 3667\\
Pions& 249 & Pions & 1.4\\
Others & 83 & Others &  $<1$\\
Total & 68163 & Total & 3714\\
\hline \hline
\end{tabular}
\end{table} 
It is also interesting to show the overall cosmogenic production in the detector. As shown in Table~\ref{tab:total800c} and Fig.~\ref{fig:total800}, the cosmogenic isotopes
induced by muons
range from P to Ca, a total rate of 19 Hz. Some of them can be background to the proposed physics channels.

\begin{table}[htb!!!]
\caption{ Additional significant cosmogenic production rates in the detector (20 kton) at the 800-ft level from the Monte Carlo simulation.}
\label{tab:total800c}
\begin{tabular}{|l|l|l|l|l|}
\hline \hline 
Isotope&Produced by&Rate per day&Q (MeV) &t$_{1/2}$\\
\hline
$^{30}$P  & Spallation & 9020 &4.23&2.5 m\\
$^{32}$P  & Spallation& 20900 & 1.71 & 14. 3 d\\
$^{33}$P  & Spallation & 30100 & 0.25 & 25.3 d\\
$^{34}$P  & Spallation & 12090 & 5.4 & 12.4 s\\
$^{35}$P  & Spallation & 7500 & 4.0 & 47. 2 s\\
$^{36}$P  & Spallation & 1190 & 10.4 & 5.6 s\\
$^{37}$P  & Spallation & 550 & 7.9 & 2.3 s\\
$^{31}$S  & Spallation & 5500 & 5.4 & 2.6 s\\
$^{35}$S  & Spallation & 215500 & 0.17 & 87.5s\\
$^{37}$S  & (n,$\alpha$) & 31500 & 4.9 & 5.1 m\\
$^{38}$S  & Spallation & 11500 & 2.9 & 170 m\\
$^{39}$S  & Spallation & 850 & 6.6 & 11.5 s\\
$^{33}$Cl & Spallation & 670 & 5.6 & 2.5 s \\
$^{34}$Cl & Spallation & 8700 & 5.6 & 32  m \\
$^{36}$Cl & Spallation& 1005000& 0.7 & 3.1$\times$10$^{5}$ y \\
$^{38}$Cl & Spallation & 110000& 4.9 & 37.24 m \\
$^{35}$Ar & (n,6n$'$) & 7100& 6.0 & 1.8 s \\
$^{37}$Ar & (n,4n$'$) & 21000& 0.8 & 35 d \\
$^{39}$Ar & (n,2n$'$) & 91000& 0.57 & 269 y \\
$^{41}$Ar &  capture & 45100& 2.5 & 109 m \\
$^{38}$K & Spallation & 650 & 5.9 & 7.6 m \\
$^{40}$K & (p,n) & 6500 & 1.3 & 1.28$\times$10$^{9}$ y \\
Total &  & 1641920&&\\
\hline \hline
\end{tabular}
\end{table} 

\subsubsection{Cosmogenic Production at the 4850-ft Level}
\begin{figure}[htb!!!]
\includegraphics[angle=0,width=8.0cm] {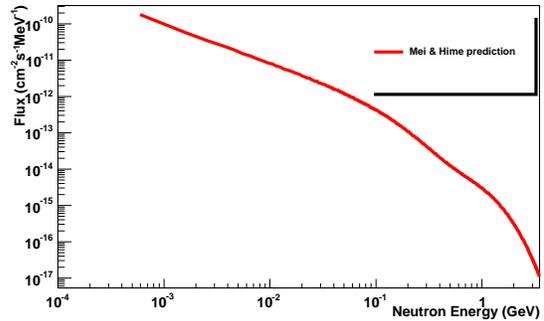}
\caption{\small{
Local fast neutron energy spectrum at the 4850-ft level used in the Monte Carlo simulation.}}
\label{fig:neuspe}
\end{figure}

The residual muons at the 4850-ft level have an average energy of 321 GeV using Eq.\ref{eq:MeiHime9}. The total flux is predicted to be
4.4$\times$10$^{-9}$cm$^{-2}$s$^{-1}$ in Ref.~\cite{meihime}.  
The neutrons that are produced in the rock by these residual high energy muons entering the experimental hall were simulated by Mei \& Hime in great detail~\cite{meihime}. 
We obtained 
a muon-induced neutron energy spectrum using the method described in Section II A. Fig.~\ref{fig:neuspe} shows the neutron energy 
spectrum. The total flux at the 4850-ft level is 5.4$\times$10$^{-10}$cm$^{-2}$s$^{-1}$ from the Monte Carlo simulation. The production of $^{40}$Cl is via (n,p) reaction on $^{40}$Ar 
requiring a threshold of 6.87 MeV. The total neutron flux with neutron energy greater than 6.87 MeV is about 1.6$\times$10$^{-10}$cm$^{-2}$s$^{-1}$, also derived from 
the Monte Carlo simulation. 

\begin{figure}[htb!!!]
\includegraphics[angle=0,width=8.0cm] {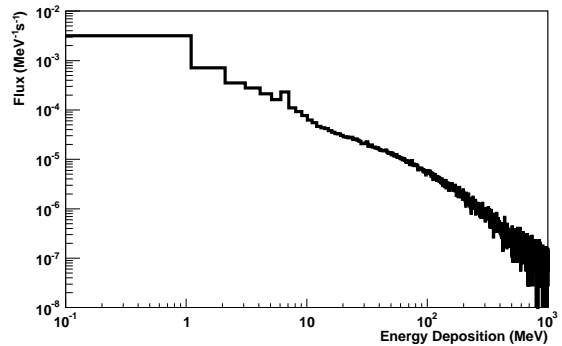}
\caption{\small{
Fast neutrons induced visible energy spectrum in the detector at the 4850-ft level. Lindhard's ionization efficiency~\cite{lindhard} was applied to the nuclear recoil events in this plot.}}
\label{fig:neuvisible4850}
\end{figure}

Similar to the simulation for the 800-ft level, we simulated both the residual muons crossing the detector and the muon-induced 
neutrons entering the detector from rock. The visible energy spectrum induced by the muon-induced neutrons entering the detector from the
surrounding rock is shown in Fig.~\ref{fig:neuvisible4850}. The event rate above 5 MeV is $\sim$0.001 Hz. The direct muon rate is $\sim$0.05 Hz.

Table~\ref{tab:g4} shows the simulated $^{40}$Cl
production rates from two sources. 

\begin{table}[htb!!!]
\caption{ $^{40}$Cl production rates in the detector (20 kton) at the 4850-ft level from the Monte Carlo simulation.}
\label{tab:g4}
\begin{tabular}{|l|l|l|l|}
\hline \hline
\multicolumn{2}{c|}{From $\mu$ simulation} & \multicolumn{2}{c|}{From $n$ simulation}\\
\hline
Produced by & Rate per day & Produced by & Rate per day\\
\hline
Muon Capture  & 17.5 & Secondary $\mu$ & 0.43\\
Secondary n & 54.4 & Neutrons & 9.3\\
Pions& 0.33 & Pions & 0.016\\
Others & 0.04 & Others &  0.002\\
Total & 72.3 & Total & 8.41\\
\hline \hline
\end{tabular}
\end{table} 

\begin{figure}[htb!!!]
\includegraphics[angle=0,width=8.0cm] {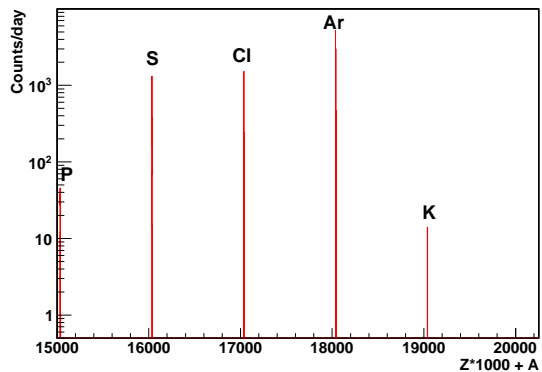}
\caption{\small{
Muon-induced cosmogenic production at the 4850-ft level from the Monte Carlo.}}
\label{fig:total4850}
\end{figure}

Similar to the above discussion of the cosmogenic production for the 800-ft level, we show overall cosmogenic isotopes induced by muons for the 4850-ft level
in Table~\ref{tab:total4850c} and Fig.~\ref{fig:total4850}. Note that the production rate is relatively small at this depth.
\begin{table}[htb!!!]
\caption{ Additional significant cosmogenic production rates in the detector (20 kton) at the 4850-ft level from the Monte Carlo simulation.}
\label{tab:total4850c}
\begin{tabular}{|l|l|l|l|l|}
\hline \hline 
Isotope&Produced by&Rate per day&Q (MeV) &t$_{1/2}$\\
\hline
$^{30}$P  & Spallation & 9.6 &4.23&2.5 m\\
$^{32}$P  & Spallation& 22.2 & 1.71 & 14. 3 d\\
$^{33}$P  & Spallation & 31.9 & 0.25 & 25.3 d\\
$^{34}$P  & Spallation & 12.8 & 5.4 & 12.4 s\\
$^{35}$P  & Spallation & 8.0 & 4.0 & 47. 2 s\\
$^{36}$P  & Spallation & 1.3 & 10.4 & 5.6 s\\
$^{37}$P  & Spallation & 0.6 & 7.9 & 2.3 s\\
$^{31}$S  & Spallation & 5.8 & 5.4 & 2.6 s\\
$^{35}$S  & Spallation & 228.5 & 0.17 & 87.5s\\
$^{37}$S  & (n,$\alpha$) & 33.4 & 4.9 & 5.1 m\\
$^{38}$S  & Spallation & 12.2 & 2.9 & 170 m\\
$^{39}$S  & Spallation & 0.9 & 6.6 & 11.5 s\\
$^{33}$Cl & Spallation & 0.7 & 5.6 & 2.5 s \\
$^{34}$Cl & Spallation & 9.2 & 5.6 & 32  m \\
$^{36}$Cl & Spallation& 1065.7& 0.7 & 3.1$\times$10$^{5}$ y \\
$^{38}$Cl & Spallation & 116.6& 4.9 & 37.24 m \\
$^{35}$Ar & (n,6n$'$) & 7.5& 6.0 & 1.8 s \\
$^{37}$Ar & (n,4n$'$) & 22.3& 0.8 & 35 d \\
$^{39}$Ar & (n,2n$'$) & 96.5& 0.57 & 269 y \\
$^{41}$Ar &  capture & 47.8& 2.5 & 109 m \\
$^{38}$K & Spallation & 0.69 & 5.9 & 7.6 m \\
$^{40}$K & (p,n) & 6.9 & 1.3 & 1.28$\times$10$^{9}$ y \\
Total & &1741&&\\
\hline \hline
\end{tabular}
\end{table} 

\subsection{Development of Analytic Models}
It is of general interest to have analytic models that can be used to estimate the cosmogenic production rates without doing a complicated simulation procedure.
Even after a campaign of Monte Carlo simulation, it is necessary to assess whether a correct result was delivered by a complicated simulation package.
 Analytic models are 
developed based on the physics processes that are then compared to the experimental results. Therefore, analytic models can be used to compare and evaluate the results 
from a full Monte Carlo simulation. We develop analytic models below.
\subsubsection{Solid Angle}
When considering the interaction of particles in a detector, it is necessary to calculate 
the solid angle subtended a certain distance from the target. Using the work of H. Gotoh 
and H. Yagi~\cite{got}, the solid angle of a point particle was calculated at an 
arbitrary distance from the detector as well as the subtended solid angle from a fixed 
point throughout the detector material. For an assumed LBNE detector of dimensions (2$w$ $\times$ 2$l$ $\times$ 
$h$) = (45.6 $\times$ 22.4 $\times$ 14) m$^{3}$, the solid angle from a point ($x_{p}$, $y_{p}$, $z_{p}$) 
is
\begin{eqnarray}
\Omega = \arctan\frac{(x_p + w)(y_p + l)}{z_p((x_p + w)^2 + (y_p + l)^2 + z_p^2)^{1/2}} \nonumber\\
 - \arctan\frac{(x_p + w)(y_p - l)}{z_p((x_p + w)^2 + (y_p - l)^2 + z_p^2)^{1/2}} \nonumber \\
 - \arctan\frac{(x_p - w)(y_p + l)}{z_p((x_p - w)^2 + (y_p + l)^2 + z_p^2)^{1/2}} \nonumber \\
 + \arctan\frac{(x_p - w)(y_p - l)}{z_p((x_p - w)^2 + (y_p - l)^2 + z_p^2)^{1/2}}. 
\end{eqnarray}
With this equation, the average solid angle for a sheet of particles generated immediately 
above the detector was calculated to be 1.74 sr, and the differential muon flux 
from~\cite{feg} could be used in the calculation of the total muon flux for a given level at the Homestake mine.

\subsubsection{Capture Rate for Negative Muons}
The stopping muons can be captured by an argon nucleus to generate radioactive isotopes such as $^{40}$Cl, etc. There are two sources of stopping muons for a given depth
in an underground laboratory: 1) the through-going muons come to the end of their energy range and 2) the secondary muons generated locally  by the primary muons and their
daughter pions. 
The capture rate for muons as a function of the depth of the detector can be calculated 
using the following equation:
\begin{equation}
R_{\mu}^{capt} = R_{\mu}^{S}\cdot f_{capt}\cdot f_c\cdot f_{ch}\cdot f_g,
\end{equation}
where the contributing terms are as follows:
\begin{enumerate}
\item{$R_{\mu}^S$ is the stopping muon rate (derived below).}
\item{$f_{capt}$ is the fraction of muons that are captured,
\begin{equation}
f_{capt} = \frac{\tau_{lifetime}}{\tau_{capt}},
\end{equation}
with $\tau_{capt}$ = muon capture time in argon, 
\begin{equation}
\frac{1}{\tau_{capt}} = \frac{1}{\tau_{lifetime}} - \frac{1}{\tau_{0}},
\end{equation}
$\tau_{lifetime}$ = 537$\pm$32 ns in argon~\cite{suzuki} and $\tau_0$ = lifetime of muons in a 
vacuum. Here, $f_{capt}$ = 0.76.}
\item{$f_c$ is the elemental fraction of the target in the compound,
\begin{equation}
f_c = \frac{a_i\cdot Z_i}{\sum_i a_i\cdot Z_i}.
\end{equation}
For the LBNE detector $f_c$ = 1.}
\item{$f_{ch}$ is the charge ratio of negative muons to total muons,
\begin{equation}
f_{ch} = \frac{\mu^-}{\mu^- + \mu^+}.
\end{equation}
It is 0.44 on the surface, but it is assumed that underground the fraction will be 
similar.}
\item{$f_g$ is the fraction of reactions that occur with $^{40}$Cl in the ground state. Two 
fractions are considered: $f_{ga}$ = 0.0712~\cite{avk} and $f_{gb}$ = 0.2 (from the Geant4 simulation).}
\end{enumerate}

The stopping muon rate can be calculated with the following formula:
\begin{equation}
R_{\mu}^S = R_{\mu}^T\cdot R\cdot f_{scale},
\end{equation}
where $R_{\mu}^T$ is the through-going muon rate, $R$ is the ratio of stopping to through-going muons, and $f_{scale}$,  the scale factor,
$\frac{mass}{area}\frac{1}{100 g\cdot cm^{-2}}$, scales from a 1 m.w.e 
detector to a larger size and was calculated to be 19.5 for the simulated LBNE detector.

The through-going muon rate is defined as
\begin{equation}
R_{\mu}^T = \phi_{\mu}^T\cdot \Omega \cdot S_{area},
\end{equation}
where $\phi_{\mu}^T$ is the differential through-going muon flux, $\Omega$ is the solid angle, and $S_{area}$ is the area through which the 
through-going muons transverse the detector. 
Using the measured differentiated through-going muon flux from F.E. Gray {\it et al.}~\cite{feg} and 
the average solid angle, the calculated through-going muon rate is 4.2$\times10^{6}$ per day at 800-ft and 5910 per day at 4850-ft.

Two equations were analyzed for the ratio of stopping to through-going muons. The first,
\begin{equation}
R_1 = \frac{0.3}{<E_{\mu}>} + 5.7\cdot 10^{-5}\cdot n_0\cdot <E_{\mu}>^{0.7},
\end{equation}
was proposed by Chudakov {\it et al.} in Ref.~\cite{chu} using experimental data, where the first term calculates the ratio of the stopping 
muons to the through-going muons for the muons from the surface and the 
second term describes the contribution of the stopping muons from the muons produced
 by cascades with $<E_{\mu}>$ defined in Eq.\ref{eq:MeiHime9} and
$n_0$ varying from 0.4-0.75 depending on depth (here, $n_0$ = 0.4 was used to fit the equation best with simulated data). 
The second equation considered,
\begin{equation}
R_2 = \gamma_{\mu}\frac{\Delta E\cdot exp(\frac{h}{\xi})}{[exp(\frac{h}{\xi})-1]\epsilon_{\mu}},
\end{equation}
is a parametrization from~\cite{tkg}, where $\gamma_{\mu}$ = 3.77~\cite{deg2}, $\xi$ = 2.5 km.w.e., $\Delta E$ $\approx$ $\alpha \Delta x$, 
$\alpha$ = 0.268 GeV/km.w.e. (for $E_{\mu}$ $\geq$ 1000 GeV) and $\Delta$x = 100 g cm$^{-2}$, $h$ is the depth 
in km.w.e, and $\epsilon_{\mu}$ = 618 GeV~\cite{tkg, deg1}.

At each depth, two rates were calculated for stopping muons and compared to the simulated 
values. These can be seen in Table~\ref{tab:Rstop}.
\begin{table}[htb!!!]
\caption{Comparison of calculated stopping muon rates to simulated results.}
\label{tab:Rstop}
\begin{tabular}{|c|c|c|}
\hline \hline
& \multicolumn{2}{c|}{Level}\\
\hline
Process & 800 & 4850\\
\hline
$R^S_{\mu,1}$ (per day) & 3.06$\times10^{5}$ & 251\\
$R^S_{\mu,2}$ (per day) & 5.45$\times10^{5}$ & 233\\ 
Sim. (per day) & 3.69$\times10^{5}$ & 173\\
\hline \hline
\end{tabular}
\end{table}

The uncertainty in the two equations above was not discussed in the references~\cite{chu,tkg}. A 30\% difference between the calculations 
and the Monte Carlo is seen in Table~\ref{tab:Rstop}. Since the result of the Monte Carlo simulation package for the muon-induced neutrons 
was compared to the surface data, we state that the stopping muon rates are within 30\% of the simulated values.

The total capture rate of negative muons was calculated using the two equations for the ratio 
of stopping to through-going muons ($R_{1}$ and $R_{2}$) and the two different ground state 
fractions ($f_{ga}$ and $f_{gb}$) resulting in four values (Table~\ref{tab:Rcapt}).
\begin{table}[htb!!!]
\caption{Comparison of calculated muon capture rates to simulated results.}
\label{tab:Rcapt}
\begin{tabular}{|c|c|c|}
\hline \hline
 & \multicolumn{2}{c|}{Level}\\
\hline
Process & 800 & 4850\\
\hline
$R^{capt}_{\mu,1a}$ (per day) & 7286 & 5.97\\
$R^{capt}_{\mu,2a}$ (per day) & 1.30$\times10^{4}$ & 5.55\\
\hline
$R^{capt}_{\mu,1b}$ (per day) & 2.05$\times10^{4}$ & 16.8\\
$R^{capt}_{\mu,2b}$ (per day) & 3.64$\times10^{4}$ & 15.6\\
\hline
Sim. (per day) & 2.73$\times10^{4}$ & 17.5\\
\hline \hline
\end{tabular}
\end{table}

\begin{figure}[htb!!!]
\includegraphics[angle=0,width=8.cm] {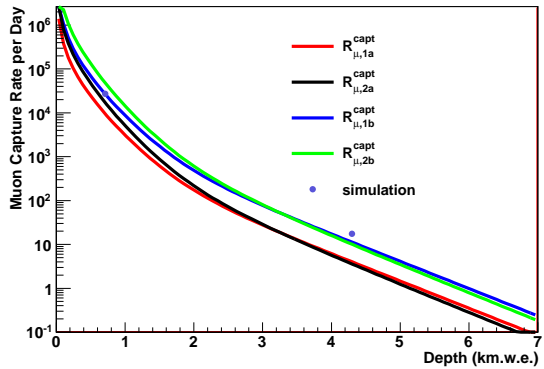}
\caption{\small{Capture rates as a function of depth. }}
\label{fig:capture}
\end{figure}

The capture rates and simulated values were plotted as a function of depth (km.w.e) in Fig.~\ref{fig:capture} using 
the integrated neutron flux from Eq.\ref{eq:MeiHime13} instead of the differential flux 
at the specific levels. As it is apparent in Fig.~\ref{fig:capture}, the use of $R^{capt}_{\mu,1b}$ 
to calculate the muon capture rate has the best agreement to Monte Carlo.

\subsubsection{Production Rate via (n,p) Reaction}
Another important reaction that contributes to the production of $^{40}$Cl in the detector 
is the (n,p) reaction, $^{40}$Ar(n,p)$^{40}$Cl. To calculate the production rate the 
following formula can be used:
\begin{equation}
\label{eq:np}
P_{(n,p)} = \frac{\Phi_n(E_n)\cdot exp(-\frac{<L>}{\lambda_{total}})[1-exp(-\frac{<L>}{\lambda_{(n,p)}})]\cdot m}{\rho\cdot <L>}
\end{equation}
where $m$ is the mass of the detector, $\rho$ is the density of the detector medium, $<L>$ is 
the average path length, $\Phi_n(E_n)$ is the integrated neutron flux, $\lambda_{total}$ is 
the mean free path considering all neutron disappearance reaction channels except (n,p), and 
$\lambda_{(n,p)}$ is the mean free path of the (n,p) reaction channel.

The first component of the (n,p) production rate, $\Phi_n(E_n)\cdot exp(-\frac{<L>}{\lambda_{total}})$, 
is the probability that neutrons will survive all reactions except the (n,p) reaction, reducing the 
flux and availability of neutrons for the production of $^{40}$Cl. Similarly, the second 
term, $[1-exp(-\frac{<L>}{\lambda_{(n,p)}})]$, is the probability that the remaining neutrons 
will undergo (n,p) reaction and produce the $^{40}$Cl background. The remaining terms, 
$\frac{m}{\rho\cdot<L>}$, are characteristics of the detector.

In calculating the average path length, the angular dependence of the neutrons (and their 
parent particle muons) was considered using the methodology of Jostlein and 
McDonald~\cite{jos}. A $\frac{1}{\cos\theta}$ dependence was used in the calculation.

The mean free path was calculated by
\begin{equation}
\lambda = \frac{A}{\rho\cdot N_a\cdot \sigma},
\end{equation}
where $A$ is the mass number of the target nucleus, $\rho$ is the density, $N_a$ is Avogadro's number, 
and $\sigma$ is the cross section. For the calculation, the flux weighted cross section, $\bar{\sigma}$, was used
\begin{equation}
\bar{\sigma} = \frac{\sum_i \phi_i\cdot \sigma_i}{\sum_i \phi_i},
\end{equation}
where flux and cross section of the same energy, $i$, are summed together. The cross sections 
from Geant4 were used for this calculation. It is important to use the flux weighted cross 
sections to calculate the mean free path in order to account for the entire spectrum of 
possible neutron energies. Unfortunately the Geant4 cross section data does not exceed 
100 MeV; however, the calculated values should be accurate to within a factor of two. This was evaluated
using the TALYS nuclear package~\cite{talys} with neutron cross section up to 250 MeV. 

Using the above formulas, we have calculated the average path length, the flux weighted
cross sections, and the mean free path for neutrons at the levels of 800-ft and 4850-ft. Table~\ref{tab:np} shows the results.
As can be seen in Table~\ref{tab:np}, the values are similar as anticipated. 
\begin{table}[htb!!!]
\caption{The calculated parameters for Eq.~\ref{eq:np}. }
\label{tab:np}
\begin{tabular}{c|c|c|c|c|c|c|}
\hline \hline
 &Path length &\multicolumn{2}{c|}{Cross section} & \multicolumn{2}{c|}{Mean free path}\\
Level&$<L>$ (cm) & $\bar{\sigma}_{(n,p)}$ (b)  & $\bar{\sigma}_{total}$ (b) & $\lambda_{(n,p)}$ (cm) & $\lambda_{total}$ (cm)\\
\hline
800 & 572.2 &0.0209 & 0.181 & 2280 & 262.9\\
4850 & 570.8 & 0.0209 & 0.181 & 2280 & 262.9\\
\hline \hline
\end{tabular}
\end{table}

Two main sources of neutrons are considered for both 800-ft and 4850-ft: 1) the muon-induced neutrons entering the detector from the
experimental hall (Source I neutrons) and 2) the muon-induced neutrons in the detector when muons pass through the target (Source II neutrons). 
The final calculated (n,p) production rate for both levels are listed in Table~\ref{tab:nprate1}  with the relevant neutron flux
in the detector above the (n,p) reaction threshold. 
\begin{table}[htb!!!]
\caption{The calculated (n,p) production rates.}
\label{tab:nprate1}
\begin{tabular}{c|c|c|c|c|}
\hline \hline
&&$\Phi(E_{n})$ (cm$^{-2}$s$^{-1}$) &\multicolumn{2}{c|}{$^{40}$Cl rate per day}\\
\hline
&Level&Flux &Analytic models &Geant4\\
Source I&800 &8.8$\times$10$^{-7}$ & 4784 & 3667\\
Neutrons& 4850 & 1.66$\times$10$^{-10}$& 9.07& 9.3\\
\hline
& Level&Flux&Analytic models & Geant4\\
Source II &800 &1.4$\times$10$^{-6}$ & 66978 & 40587 \\
Neutrons& 4850 & 6.0$\times$10$^{-10}$ & 54& 54\\
\hline \hline
\end{tabular}
\end{table}

As shown in Table~\ref{tab:nprate1}, the predicted production rates using the analytic models agree with
the Geant4 simulation within 30\% for the 800-ft level and less than 1\% for the 4850-ft level. This is because
the angular dependence of neutrons, $\frac{1}{\cos\theta}$, used in the analytic models works better at large depth~\cite{meihime}.
Lack of proper neutron angular distribution for shallow depths results in a slightly large discrepancy
in the production rates between the analytic models and the Geant4 simulation.  
Note that the analytic calculation offers only a crosschecking. The agreement between the 
Monte Carlo simulation and the analytic calculation indicates that similar physics processes are implemented. 

\subsection{Scaling Function}
Muon-induced processes and
the cosmogenic radioactivity production  depends strongly on the target and must be 
evaluated individually for the experiment. However, the production rate is proportional
to muon flux, or neutron flux, and their interaction cross-section.  The energy dependence
of the total cross-section for all muon-induced radio-isotopes in the scintillator
was evaluated assuming the power law~\cite{ffk}
\begin{equation}
\label{eq:cross1}
\sigma_{tot}(E_{\mu}) \propto E_{\mu}^{\alpha},
\end{equation}
where $\alpha$ varies from 0.50 to 0.93 with a weight mean value $<\alpha>$ = 0.73$\pm$0.10
~\cite{hag}. For a given number of target atoms $N$ and the cross-section $\sigma_{0}$ at the Earth
surface where the average muon energy is about 4 GeV, the muon-induced cosmogenic radioactivity ($R_{iso}$)
depends on the differential muon energy spectrum $dN_{\mu}$/$dE_{\mu}$ at the experimental site
at a depth $h_{0}$,
\begin{equation}
\label{eq:cross2}
R_{iso} = N\sigma_{0}\int_{0}^{\infty} \left(\frac{E_{\mu}}{4\; GeV}\right)^{\alpha} \frac{dN_{\mu}}{dE_{\mu}}dE_{\mu}.
\end{equation}
As a simplification, the production rate is written as a function of the average
muon energy $<E_{\mu}>$ at a depth $h_{0}$~\cite{hag}:
\begin{equation}
\label{eq:cross3}
R_{iso} = \beta_{\alpha}N\sigma_{4\; GeV}\left(\frac{<E_{\mu}>}{4\; GeV}\right)^{\alpha}\phi_{\mu},
\end{equation}
where $\phi_{\mu}$ is the total muon flux at the experimental site and $\beta_{0.73}$ = 0.87 $\pm$ 0.03
is the correction factor for the averaging of $E_{\mu}$~\cite{hag}. For a given detector target and a depth,
the cosmogenic production rate as a function of depth is thus obtained
\begin{equation}
\label{eq:cross4}
\frac{R_{iso}(unknown)}{R_{iso}(known)} = \left(\frac{E_{\mu,unknown}}{<E_{\mu, known}>}\right)^{\alpha}\frac{\phi_{\mu}(unknown)}{\phi_{\mu}(known)}.
\end{equation}

\section{Depth-Sensitivity Relation}
LBNE is an extremely rich physics program that will measure neutrino properties using a neutrino beam. In addition, LBNE will also measure supernova neutrinos and proton decays. 
Each of these physics channels has its unique signal region in terms of energy distribution. For example, measuring the parameters of neutrino oscillation with a neutrino beam 
has an energy region from 1 to 8 GeV while the energy region of proton decay ranges from 100 to 938 MeV. The signal of supernova neutrinos resides between 
5 to 50 MeV. It is
difficult to establish a depth-sensitivity relation for all physics channels using a single plot. We elaborate on the muon-induced backgrounds for each of the physics channels below.

\subsection{Muon and the Muon-induced Neutron Rates}
\begin{figure}[htb!!!]
\includegraphics[angle=0,width=8.cm] {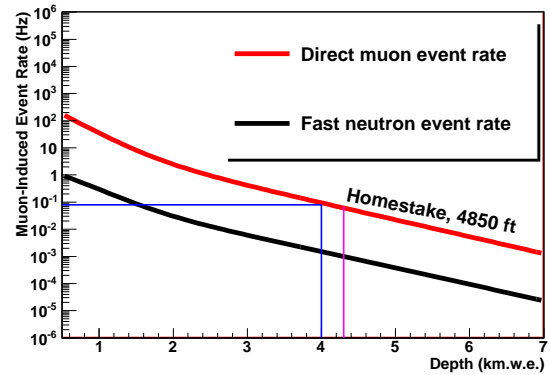}
\caption{\small{Muon and muon-induced neutron rates as a function of depth. Shown is for the energy deposition greater than 5 MeV in the detector.}}
\label{fig:neurate}
\end{figure}

Utilizing the above formulas, the muon and the muon-induced neutron rates as a function of depth can be shown in Fig.~\ref{fig:neurate}.
It is clear that the event rates decrease rapidly when the depth increases. 
The fluctuation of muon and the muon-induced neutron rates from seasonal variation can result in backgrounds 
for all
physics channels depending on the depth.  
\subsection{Coincidence Rate with the Beam Physics}
\begin{figure}[htb!!!]
\includegraphics[angle=0,width=8.cm] {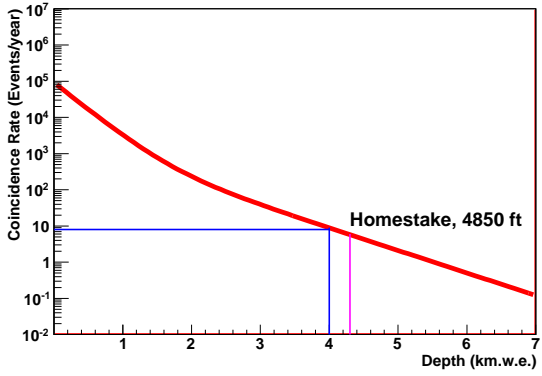}
\caption{\small{The coincidence rate of muons and the muon-induced processes with neutrino beam spills is shown as a function of depth.}}
\label{fig:coin}
\end{figure}

The LBNE experiment intends to use an accelerator with a cycle time of 1.33 seconds for a proton pulse of 10 $\mu$s~\cite{lbnecdr}. Therefore, the lifetime of a neutrino beam is
about 237 seconds per year.  
The relevant muon-induced processes that are backgrounds to the beam physics are the events falling in the energy region of 1 to 8 GeV within the drift time of the detector for neutrino beam events. 
The coincidence rate of muons and the muon-induced processes with the neutrino beam spills as a function of depth is calculated assuming a drift time of 2 ms. Fig.~\ref{fig:coin} 
displays the result. It is clear that a rejection power of ~10$^{5}$ is needed if the detector is built on the surface with a depth of 5 m.w.e. Note that the coincidence rate shown in
Fig.~\ref{fig:coin} does not apply any rejection power. 

\subsection{Cosmogenic Production Rates}
\begin{figure}[htb!!!]
\includegraphics[angle=0,width=8.cm] {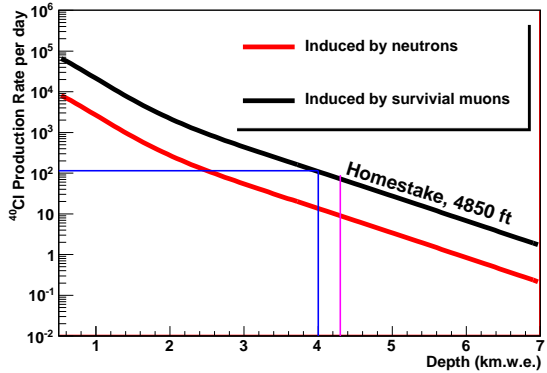}
\caption{\small{$^{40}$Cl production rates produced by (n,p) reaction as a function of depth. }}
\label{fig:cross5}
\end{figure}

The cosmogenic production rate also decreases with increasing depth.
As an example, Fig.~\ref{fig:cross5} shows the $^{40}$Cl production rate as a function of depth. The cosmogenic produced radioactive isotopes can be backgrounds for the detection of 
relic supernova neutrinos.
\subsection{$\pi_{0}$ Production Rates}
\begin{figure}[htb!!!]
\includegraphics[angle=0,width=8.cm] {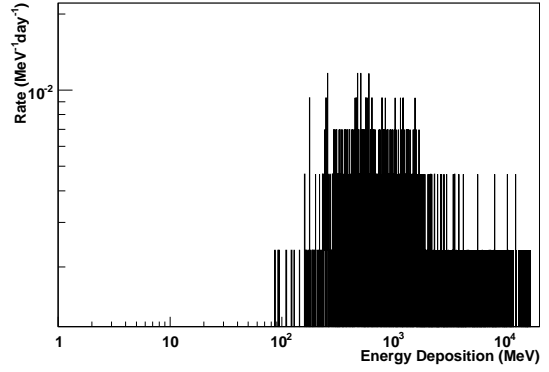}
\caption{\small{$\pi_{0}$ events in the detector at the 4850-ft level. }}
\label{fig:pion0k}
\end{figure}

\begin{figure}[htb!!!]
\includegraphics[angle=0,width=8.cm] {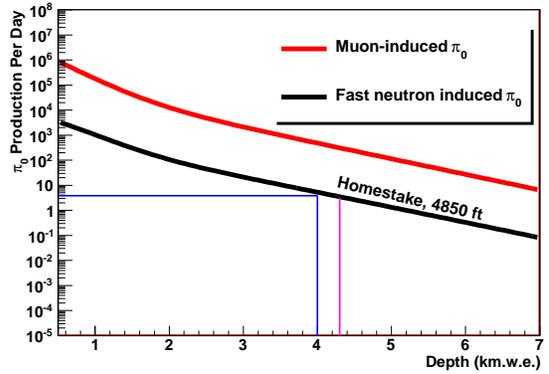}
\caption{\small{$\pi_{0}$ production rates as a function of depth. }}
\label{fig:pion0}
\end{figure}

The neutrino beam-induced neutral current and charge current $\pi_{0}$ productions are important backgrounds to the $\nu_{e}$ appearance. However, $\pi_{0}$ can also be produced by muons and muon-induced neutrons in the detector.
Fig.~\ref{fig:pion0k} shows the energy deposition from $\pi_{0}$ events created by fast neutrons in the detector. It is worth mentioning that there are also multiple $\pi_{0}$
events along the neutron track.
The production of $\pi_{0}$ as a function of depth is shown in Fig.~\ref{fig:pion0}. It is clear that the $\pi_{0}$ produced by fast neutrons entering the detector can be 
a significant background to the beam physics program. This is because the fluctuation of the $\pi_{0}$ production due to the seasonal variation of muon flux and neutron flux results in
a level of 22 $\pi_{0}$ events per year in the detector. In addition, the statistical fluctuation of the $\pi_{0}$ has a similar level of 33 $\pi_{0}$ events. Adding both in quadratic,
the level
of fluctuation in the production of $\pi_{0}$ can be about 40 events per year. This is significant even at the 4850-ft level. 
Depending on the capability of discriminating single $\pi_{0}$ events between the beam neutrino-induced and the cosmic neutron-induced, a greater depth (7400-ft level) 
can be an effective option to further reduce this background.

\subsection{Backgrounds for $\nu_{e}$ Appearance}
\begin{figure}[htb!!!]
\includegraphics[angle=0,width=8.cm] {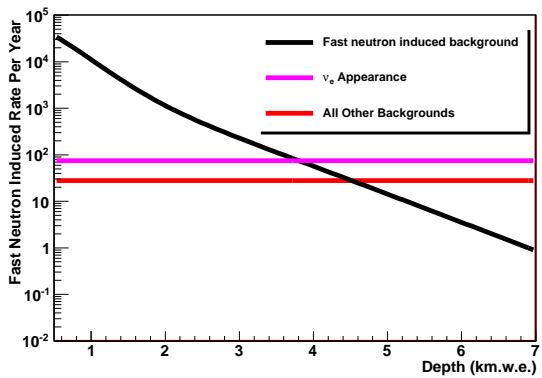}
\caption{\small{Muon-induced background as a function of depth. Shown is the average value of statistical fluctuation induced by fast neutrons. 
Note that signal and the other total backgrounds were estimated using Ref.~\cite{beam}. }}
\label{fig:back}
\end{figure}

The $\nu_{e}$ appearance is essential to the measurements of neutrino properties with a neutrino beam. The anticipated signal in a 20 kton 
LAr detector is about 75 events per year~\cite{beam}. Muons and the muon-induced processes can generate $\nu_{e}$-like events 
in the detector through the following:
1) the production of energetic delta electrons in the ionization process, which 
are very hard to reject using the reconstruction of muon tracks, because a fraction of them do 
not even have the parent muon tracks associated with them;
2) muon-induced  bremsstrahlung radiation, pair production, and $\pi_{0}$;
3) high energy neutrons from the surrounding materials; and
4) high-energy gamma rays 
produced by muon bremsstrahlung radiation in the surrounding materials.
The muon-induced backgrounds can be measured with the beam-off.
 However, the fluctuation of background events resulting from statistical and seasonal 
variation is a main source of background.
 This background as a function of depth is shown in Fig.~\ref{fig:back}. As can be seen in Fig.~\ref{fig:back}, the depth must be greater than 4.0 km.w.e. in order to
have a reasonable measurement of $\nu_{e}$ appearance at SURF.

\section{Conclusion}
We evaluate the muon-induced background as a function of depth for a long baseline neutrino experiment with liquid argon as the target at different levels  
of the Sanford Underground Research Facility at Homestake mine. 
Both Geant4 simulations and analytic methods are employed in the evaluation of background event rates in the region of interest
 using the available muon and neutron energy spectra
 from Ref.~\cite{meihime} and 
the measured muon flux from Ref.~\cite{feg}. The muon and muon-induced neutron rates are calculated for the energy greater than 5 MeV. The production of 
energetic delta electrons in the ionization process is found to be the most important
 muon-induced $\nu_{e}$-like events for the beam physics.
The cosmogenic processes
are discussed in detail for negative muon capture, (n,p), and (p,n) reactions. The dominant backgrounds are from both stopping muons and the high energy neutrons 
generated by muons in the surrounding materials and in the target. In summary, the background sources are:
1) the negative stopping muon capture; 2) the neutron (n,p), (n,d),(n,t), (n,$\alpha$), etc; and 3) the muon-induced energetic delta electrons and showers, in particular 
high energy delta electrons, $\pi_{0}$ production, and high 
energy gamma rays.  
The cosmogenic production rate as a function of depth is evaluated for negative stopping muon capture and fast neutrons separately.
As can be seen from the above discussion, the cosmogenic production rate induced by the muon-induced processes reduces three orders of magnitude
 when the depth is larger than 4.0 km.w.e. 
It is clear from Fig.~\ref{fig:capture} and Fig.~\ref{fig:cross5} that the $^{40}$Cl production is less than 100 per day when the depth is greater than 4.0 km.w.e.  
We conclude the following: 
\begin{itemize}
\item{The 800-ft level presents large background events for beam-related neutrino physics
because the muon rate ($\sim$88 Hz) is still high and the $\nu_{e}$-like events 
produced by the muon-induced energetic
delta electrons are in the order of a few thousand per year. Though LAr TPC is better in identifying $\pi_{0}$ events
compared to a water Cerenkov detector~\cite{bob}, the $\pi_{0}$ produced by muons seen in Fig.~\ref{fig:pion0} is 
three orders of magnitude higher than the expected signal. A rejection power of 1000 is needed, which is contingent on the capability of discriminating single $\pi_{0}$ events 
between the beam neutrino-induced and the cosmic neutron-induced. An alternative approach is a greater depth (7400-ft level) 
to further reduce this background. }
\item{The 800-ft level possesses difficulty in detecting supernova neutrinos for a galactic supernova neutrino burst with a time window of 30 seconds 
at 10 kpc. This is because the expected charge current events in a 20 kton detector used in the simulation from such a burst are about 1300 events~\cite{icarus}. 
They correspond to about 44 Hz. However,
the total muon-induced rate is about 88 Hz at this level. The signal is immersed in background. In addition, the frequency of galactic supernovae occurs at a level of once 
per 50 years~\cite{nova}.
}
\item{ The muon-induced processes
are backgrounds for an argon-based detector in the detection of relic supernova neutrinos with a depth less than 4.0 km.w.e. With a 20 kton detector, we expect
less than 30 events per year from relic supernova neutrinos~\cite{relic}. The relic supernova neutrinos can be detected with the accumulation of the detector lifetime. }
\item{ Fig.~\ref{fig:back} shows a depth requirement for the $\nu_{e}$ appearance from a neutrino beam. It is clear that a meaningful measurement of CP violation can only be accomplished
when the depth is greater than 4.0 km.w.e.}
\item{ Finally, positioning a LAr detector near the surface, i.e. at NOVA depths, will increase the backgrounds by three orders of magnitude, 
compromising the direct $\nu$ program. Fig.~\ref{fig:nova} shows a simulated result for the NOVA depth. As can be seen in 
Fig.~\ref{fig:nova}, the signal from a supernova burst is completely immersed by the neutron induced background.
\begin{figure}[htb!!!]
\includegraphics[angle=0,width=8.cm] {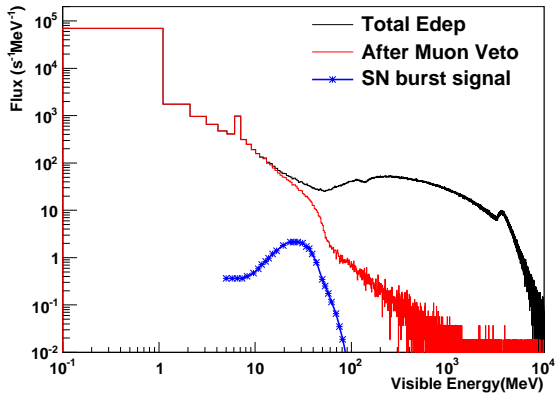}
\caption{\small{Muon-induced background for the NOVA depth. Shown is the energy deposition by muons and neutrons. }}
\label{fig:nova}
\end{figure}
 }
\end{itemize}

 Therefore, a depth larger than 4.0 km.w.e. is needed for an argon-based detector. The 4850-ft level at Homestake mine would be a good home for this detector.
 
\section{Acknowledgement}
The authors wish to thank Kevin Lesko, Bob Svoboda, Kate Scholberg, Christina Keller, and Angela A. Chiller for a careful reading of this manuscript. In particular, the authors 
would like to thank Kevin Lesko for his
many invaluable suggestions in presenting various backgrounds clearly in this paper.  
This work was supported in part by NSF PHY-0758120, DOE grant DE-FG02-10ER46709, the Office of Research at University 
of South Dakota and a 2010 research center support by the State of South Dakota. 

%
%

\begin{thebibliography}{99}
\bibitem{superk} Fukuda Y. {\it et al.} (Super-Kamiokande Collab.), Phys. Rev. Lett. 1998. V. 81. P. 1562.
\bibitem{macro} Ambrosio M. {\it et al.} (MACRO Collab.), Phys. Lett. B. 2003. V. 566. P. 35.
\bibitem{soudanii} Sanchez M. C. {\it et al.} (Soudan 2 Collab.), Phys. Rev. D. 2003. V. 68. P. 113004.
\bibitem{k2k} Aliu E. {\it et al.} (K2K Collab.), Phys. Rev. Lett. 2005. V. 94. P. 081802.
\bibitem{minos}  Michael D. G. {\it et al.} (MINOS Collab.), Phys. Rev. Lett. 2006. V. 97. P. 191801.
\bibitem{sno} Ahmad Q. R. {\it et al.} (SNO Collab.), Phys. Rev. Lett. 2002. V. 89. P. 011301.
\bibitem{homestake} Cleveland B. T. {\it et al.} Astrophys. J. 1998. V. 496. P. 505.
\bibitem{kamland} Hirata K. S. {\it et al.} (KAMIOKANDE-II Collab.), Phys. Rev. Lett. 1989. V. 63. P. 16.
\bibitem{doublechooz2} T. Matsubara Double Chooz Collab., arXiv:1205.6685v1.
\bibitem{dayabay} F.P. An {\it et al.} (Daya Bay Collab.), arXiv:1203.1669v2.
\bibitem{reno} J.K. Ahn {\it et. al.} (Reno Collab.), arXiv:1204.0626v2.
\bibitem{Barger} V. Barger {\it et al.}, Phys. Rev. D 74, 073004 (2006).
\bibitem{minos1} P. Adamson {\it et al.}, MINOS Collaboration, arXiv:1202.2772. 
\bibitem{t2k} K. Abe {\it et al.}, arXiv:1106.2822v2.
\bibitem{FMY} Fukugita M. and Yanagida T., Phys. Lett. B. 1986. V. 174. P. 45; Anisimov A., Blanchet S.,and  Di Bari P., JCAP. 2008. V. 0804. P. 033.
\bibitem{FFS} Feruglio F., Strumia A., and Vissani F., Nucl. Phys. B. 2002. V. 637. P. 345 (Nucl. Phys. B. Addendum. 2003. V. 659. P. 359).
\bibitem{lbnecdr} The LBNE Collaboration. LBNE Conceptual Design Report, Volume 5: Liquid Argon Detector for LBNE, Feb. 17, 2012.
\bibitem{LBNE}  M. Bass {\it et al.}, LBNE Collaboration, A Study of the Physics Potential of the Long-Baseline Neutrino Experiment Project
with an Extensive Set of Beam, Near Detector and Far Detector Configurations, LBNE-PWG-002, INT-PUB-11-002. V. Barger {\it et al.}, Report of the US long baseline neutrino experiment study, FERMILAB-0801-AD-E, BNL-77973-2007-IR, May 2007.
\bibitem{geant4} S. Agostinelli {\it et al.}, Nucl. Instr. and Meth. in Physics Research A 506 (2003) 250-303. J. Allison {\it et al.}, IEEE Transactions on Nuclear Science 53 No. 1 (2006) 270-278.
\bibitem{meihime}D.-M. Mei and A.Hime, Phys. Rev. D. 73 (2006) 053004.
\bibitem{tkg} Thomas K. Gaisser, {\it Cosmic Rays and Particle Physics} (Cambridge University Press, New York, 1990), p. 71.
\bibitem{sei} S. Eidelman {\it et al.} (Particle Data Group), Phys. Lett. C {\bf 15}, 1 (2000)
\bibitem{deg2} Donald E. Groom {\it et al.}, At. Data Nucl. Data Tables {\bf 78}, 183 (2001).
\bibitem{yfw} Y.-F. Wang {\it et al.}, Phys. Rev. D {\bf 64}, 013012 (2001).
\bibitem{gor} M.S. Gordon {\it et al.}, IEEE Trans. Nucl. Sci. 51(6) (2004)3427.
\bibitem{lip} P. Lipari and T. Stanev, Phys. Rev. D {\bf 44}, 3543 (1991).
\bibitem{icarus} A. Bueno, I. Gil-Botella, A. Rubbia, hep-ph/0307222v1.
\bibitem{beam} Jon Urheim, Indiana University for the LBNE Science Collaboration, Status of the LBNE Long Baseline Neutrino Experiment, Meetings of the Division of Particles 
and Fields of teh American Physical Society,  9 August 2011.
\bibitem{lindhard} J. Lindhard {\it et. al.}, Mat. Fys. Medd. K. Dan. Vidensk. Selsk. {\bf 33}, 1 (1963).
\bibitem{got} H. Gotoh and H. Yagi, Nucl. Instr. and Meth. 96 (1971) 485-486.
\bibitem{feg} F.E. Gray {\it et al.}, Nucl. Instr. and Meth. A 638 (2011) 63-66.
\bibitem{suzuki} T. Suzuki, D.F. Measday, and J.P> Roalsving, Phys. Rev. {\bf C} 35 (1987)2212.
\bibitem{avk} A.V. Kinskikh {\it et al.} Russ. Acd. Sci (2008) 72 6 735.
\bibitem{chu} Alekseev, E. N., Chudakov, A. E., Gurentsov, V. A., Mikheev, S. P., Tizengausen, V. A., Proceedings of the 13th International Conference on Cosmic Rays, held in Denver, Colorado, Volume 3 (MN and HE Sessions)., p.1936.
\bibitem{deg1} D.E. Groom {\it et al.} (Particle Data Group), Eur. Phys. J. C. {\bf 15}, 1 (2000).
\bibitem{jos} Hans Jostlein and Kirk T. McDonald {\it Path Length of Muons Traversing an Arbitrary Volume} (2007).
\bibitem{talys} A. J. Koning, S. Hilaire and M. C. Duijvestijn, “TALYS: Comprehensive nuclear
reaction modeling,” Proceedings of the International Conference on Nuclear
Data for Science and Technology - ND2004, AIP vol. 769, eds. R. C. Haight, M.
B. Chadwick, T. Kawano, and P. Talou, Sep.26-Oct. 1, 2004, Sante Fe, USA,
2005, pp. 1154.
\bibitem{ffk} F.F. Khalchukov, {\it et al.}, II Nuovo Cimento, {\bf 18 C}(5) (1995) 517.
\bibitem{hag} T. Hagner  {\it et al.}, Astroparticle Physics {\bf 14} (2000)33-47.
\bibitem{bob} Bob Svoboda, Overview of the Long Baseline Neutrino Experiment, INT Program 10-2b, Long-Baseline Neutrino Physics and Astrophysics, July 26-August 27, 2010.
\bibitem{nova} R. Diehl and C. Winkler, "Integral identifies supernova rate for Milky Way". European Space Agency. 2006-01-04. Retrieved 2007-02-02.
\bibitem{relic} A.G. Cocco, {\it et al.}, JCAP12(2004)002 doi:10.1088/1475-7516/2004/12/002.
\end{thebibliography}

\end{document}